\documentstyle[11pt,aaspp4]{article}

\begin{document}

\title{The Flatness of Mass-to-Light Ratio on Large Scales}

\author{Li-Zhi Fang\footnote{Department of Physics, University of Arizona,
Tucson, AZ 85721}, and
Wen Xu\footnote{Department of Physics and Astronomy, Arizona
State University, Tempe, AZ 85287}
}
 
\begin{abstract}

It has been suggested that the mass-to-light ($M/L$) ratio of gravitationally
clustering objects is scale-independent on scales beyond galaxy clusters, and
may also be independent of the mass of the objects. In this paper, we show 
that the scale behavior of $M/L$ ratio is closely related to the scaling of 
cosmic structures larger than clusters. The scale dependence of the $M/L$ ratio
can be determined by comparing the observed scaling of richness function
(RF) of multi-scale identified objects with the model-predicted scaling of mass 
function (MF) of large scale structures. Using the multi-scale identified 
clusters from IRAS 1.2 Jy galaxy survey, we have made comparisons of the 
observed RF scaling of IRAS $r_{cl}$-clusters with the MF scalings given 
by simulations of three popular models SCDM, LCDM and OCDM. We find that, the 
M/L ratio basically is scale-independent from the Abell radius up to about 
24 $h^{-1}$Mpc, while it seems to show a slight, but systematical, increase
over this scale range. This result is weakly dependent on the cosmological 
parameters.

\end{abstract}


\section{Introduction}

The scale-dependence of mass-to-light $M/L$ ratio is always an important
topic of cosmology. The $M/L$ ratio on scales of galaxies, double
galaxies, groups of galaxies, and galaxy clusters provided the first
approach of evaluating the mean mass density of the universe (e.g. Faber
\& Gallagher 1979). A most striking result in this study is that $M/L$
ratio increases with the scale of the system considered (e.g. Rubin 1993).
This has inspired the searching for dark matters in objects on scales larger
than clusters. Yet, the $M/L$ ratio appears to be flat on the scales
of rich clusters and larger. This result has been employed to propose that 
the $M/L$ ratio on scales larger than clusters asymptotically reaches a 
constant, which is approximately equal to the median $M/L$ ratio of rich 
clusters, i.e. $M/L \simeq 300 \pm 100 $ $h$ $(M/L)_{\odot}$ (Bahcall, 
Lubin \& Dorman 1995). This asymptotic $M/L$ ratio yields 
$\Omega_{m} \simeq 0.2 \pm 0.07$,
suggesting a low density universe (Bahcall \& Fan 1998). That is, the
behavior of $M/L$ ratio on large scales is directly related to the
cosmological parameter $\Omega_m$.

The evidence for a flat $M/L$ on large scale is still weak.
The observed Virgo cluster infall motion seems to be the only evidence of the 
flatness on the scale of the local supercluster ($\sim$ 10 $h^{-1}$ Mpc)
(e.g. Yahil, Sandage \& Tammann, 1980). However, a direct measure of the
mass-to-light ratio of the Corona Borealis supercluster gives 
$M/L \simeq 560\ h (M/L)_{\odot}$. Namely, the $M/L$ ratio on scale 
$\sim$ 20 $h^{-1}$ Mpc is higher than that of rich clusters by a factor 
of 2 (Small et al. 1998). On the other hand, galaxy clusters with Abell
radius generally are dominated by elliptical galaxies at their
centers, while larger
structures have a higher fraction of spiral galaxies. Therefore, it is  
uncertain if the morphological segregation leads to a variation of the 
$M/L$ ratio of structures larger than clusters. 

In this {\it Letter}, we try to approach this problem by analyzing the 
observed scaling of clustering on scales beyond clusters. We have found 
recently that the richness function (RF) and the mass function (MF) of objects 
identified by multi-resolution analysis are scale invariant from the Abell 
radius to about 20 $h^{-1}$ Mpc (Xu, Fang and Wu 1998, hereafter XFW). We 
will show that the scale dependence of $M/L$ ratio is completely determined 
by the scaling factors of RF and MF of these multi-scale identified 
objects. If the mass field traced by the observed galaxies can match with 
popular models of CDM cosmogony, one can calculate the $M/L$'s scale dependence 
by comparing the observed RF scaling with the model-predicted MF scaling. 
Using the data of the IRAS 1.2 Jy galaxy survey, we find that the $M/L$ ratio 
is approximately flat on scales up to, at least, 24 $h^{-1}$ Mpc.

In \S 2 we calculate the mass functions of multi-scale identified clusters
from simulation and calculate their scaling indices. In \S3 we derive the
richness functions and their scaling indices of multi-scale identified clusters
from IRAS 1.2 Jy galaxy survey. In \S 4 we study the scale dependence of 
mass-to-light ratio by matching the scaling behaviors of theoretical mass
functions with those of observational richness functions. \S 5 is our
conclusion.

\section{Mass function of $r_{cl}$-clusters from simulation samples}

The model samples of mass distribution used in this paper are similar to
XFW. They are generated by N-body simulations with the P$^3$M code developed 
by Y.P. Jing (Jing \& Fang 1994). Three popular models of the CDM cosmogony, 
i.e. the standard CDM (SCDM), the flat low-density CDM (LCDM) models and the 
open CDM model (OCDM), are employed. The Hubble constant $h$ 
($\equiv H_0/(100$ km/sec/Mpc)), mass density $\Omega_0$, cosmological 
constant $\Omega_{\lambda}$ and the normalization of the power spectrum of 
the density perturbation $\sigma_8$ are chosen to be (0.5, 1.0, 0.0, 0.62) 
for SCDM, (0.75, 0.3, 0.7, 1.0) for LCDM, and (0.75, 0.3, 0.0, 1.0) for OCDM. 
The simulation parameters are: 1.) box size $L=310\ h^{-1}$ Mpc; 2.) number 
of simulation particles $N_p=64^3$; and 3.) the effective force resolution 
$\eta=0.24\ h^{-1}$ Mpc. In this case, each particle has a mass of 
$3.14\times10^{13}\Omega_M h^{-1}M_{\odot}$, which is small enough to
resolve reliably objects with mass $M\geq 5.5 \times 10^{14} h^{-1} M_{\odot}$.We produce 8 realizations for each model.

For the simulation results, the clusters on scale $r_{cl}$ and with mass 
larger than $M$ are identified by the multi-resolution (or multi-scale) 
analysis based on discrete wavelet transformation (DWT) (Fang \& Thews 1998). 
It has been shown that the DWT-identified clusters on the Abell scale, i.e. 
$r_{ab}$-clusters, are the same as those given by {\it friends-of-friends} 
(FOF) identification (XFW; Jing \& Fang 1994). That is, in terms of the 
$r_{ab}$-cluster identification, the DWT method is as good as the FOF. 
Thus, the DWT method can be reliably used to systematically identify objects 
on scales $r_{cl} > r_{ab}$ (Xu, Fang \& Deng 1999).

With the multi-scale identification of simulation samples, we have
model-predicted the mass function of $r_{cl}$-clusters, $n(>M,r_{cl})$,
which is the number density of $r_{cl}$-clusters with masses larger than $M$. 
The MFs of $r_{cl}$-clusters with 
$r_{cl}=1.5$, 3, 6, 12 and 24 $h^{-1}$ Mpc in models of the SCDM, LCDM 
and OCDM are shown in Fig. 1. The middle part of mass function,
roughly in the range $10^{-4} > n(>M, r_{cl}) > 10^{-7}$ $h^{3}$ Mpc$^{-3}$, 
is more reliable, as all the Poisson errors are relatively small. 

It can be seen from Fig. 1  that all the MFs have a similar shape. This is 
because of the MF scale invariance (XFW) as
\begin{equation}
n(>M, r_{cl})= n(> \lambda M, r_{ab}),
\end{equation}
where the scaling factor $\lambda(r_{cl}/r_{ab})$ is only a function of
the ratio $r_{cl}/r_{ab}$, but independent of $M$. Obviously 
$\lambda=1$ when $r_{cl}=r_{ab}$. The scaling factor $\lambda$ is an 
important statistic description of clustering on large scales.

\section{Richness function of $r_{cl}$-clusters from IRAS sample}

We apply the DWT multi-resolution identification to the redshift 
survey of the Infrared Astronomical Satellite (IRAS) galaxies with flux 
limit of $f_{60}\ge1.2Jy$ (Fisher et al. 1995). The IRAS sample is of 2 
(physical space) + 1 (redshift space) dimension. A DWT algorithm for 
identifying $r_{cl}$-clusters from the 2+1 dimension samples has been 
developed (XFW; Xu, Fang \& Deng 1999). To minimize the radial effects, the 
data are divided into redshift intervals of $[2500, 5000]$, 
$[5000, 7500]$ and $[7500, 10000]$ km s$^{-1}$. We then identify 
$r_{cl}$-clusters in each interval on scales 
$r_{cl} = \ 1.5, \ 3, \ 6, \ 12,$  and $24$ $h^{-1}$ Mpc. 

The sampling rate of the IRAS sample is low. The IRAS clusters identified 
on Abell scale $r_{ab}$ contain much fewer galaxies than optical clusters, 
and IRAS galaxies do not populate the clusters
in the same manner as optical galaxies
(Fisher et al. 1995; Webster, Lahav \& Fisher 1997). This gives rise to 
higher Poisson errors. However, if we are interested in
clusters larger than the Abell 
scale, this problem would be softened as structures contain more member 
galaxies. Meanwhile, IRAS galaxies are mostly of late-type. It traces  
clusters on various scales $r_{cl}$ uniformly. Therefore, the samples of IRAS
$r_{cl}$-clusters may be less affected by morphological segregation, and 
good for studying the scale-dependence of structures.

With the identified $r_{cl}$-clusters of IRAS galaxies, one can find the
richness function of IRAS $r_{cl}$-clusters, $N_{IR}(>N_g, r_{cl}, z)$, 
which is the number density of $r_{cl}$-clusters containing $\geq N_g$ 
members of IRAS galaxies in the redshift shell $z$. The RFs 
of $r_{cl}$-clusters with $r_{cl}=1.5$, 3, 6, 12 and 24 $h^{-1}$ Mpc are 
shown in Fig. 2, in which the errors are Poissonian (Xu, Fang \& Deng 1999).
Despite the errors are large, it is clear that the shapes of all 
the RFs are basically the same. This feature can be described as
a RF scaling invariance in each $z$ shell, 
\begin{equation}
N_{IR}(>N_g,r_{cl},z)=N_{IR}(>\lambda_{IR}N_g, r_{ab},z).
\end{equation}
The scaling factor $\lambda_{IR}(r_{cl}/r_{ab})$ is a function of the scale 
ratio $r_{cl}/r_{ab}$, but independent of $N_g$. Obviously $\lambda_{IR}=1$ 
when $r_{cl} = r_{ab}$. The scaling factor $\lambda_{IR}$ of IRAS 
$r_{cl}$-clusters as a function of $r_{cl}/r_{ab}$ is presented in Fig. 3.
It shows that $\lambda_{IR}$ is also independent of $z$ within the
1-$\sigma$ error bars.

\section{Scale dependence of mass-to-light ratio}

By definition, the mass-to-luminosity ratio for IRAS $r_{cl}$-clusters is the
ratio between mass and luminosity of $r_{cl}$-clusters. One can 
introduce the $M/L$ ratio of the IRAS $r_{cl}$-clusters by a mass-number 
relation as
\begin{equation}
M= R(r_{cl})N_G,
\end{equation} 
where $M$ is the total mass of a $r_{cl}$-cluster, and $N_G$ the total number 
of member galaxies of the $r_{cl}$-cluster, defined by 
\begin{equation}
N_G= N_g/\phi(z),
\end{equation}
where $\phi(z)$ is the selection function (Fisher et al. 1995). 
The definition of $R(r_{cl})$ by eq.(3) has implicitly assumed that $R(r_{cl})$, or the $M/L$ 
ratio, is independent of the mass of the objects,  since
the $M/L$ ratio of $r_{ab}$-clusters is found to be mass-independent on average (Bahcall
\& Cen 1993). Such independence may be
caused by the fact that the scale of luminosity-related 
hydrodynamical process are smaller than that of clusters. Thus, luminosity 
functions and $M/L$ ratios  are statistically independent of 
environment on cluster scale. Therefore, it would be reasonable to assume 
that the mass-independence of the $M/L$ ratio still hold on scales larger than 
clusters. 

The number-mass conversion coefficient $R(r_{cl})$ is proportional to the 
mass-to-light ratio $M/L(r_{cl})$ for IRAS galaxies on scale $r_{cl}$. We 
have then
\begin{equation}
\frac {M/L(r_{cl})}{M/L(r_{ab})} = \frac {R(r_{cl})}{R(r_{ab})}.
\end{equation}

Using eqs. (3) and (4) , one can find the mass function of IRAS 
$r_{cl}$-clusters $n_{IR}(>M,r_{cl},z)$ as 
\begin{equation}
n_{IR}(>M,r_{cl},z) = N_{IR}(>(M\phi(z)/R(r_{cl})),r_{cl},z).
\end{equation}
Therefore, the RF scaling (2) will further yield a MF scaling of IRAS 
$r_{cl}$-clusters as
\begin{equation}
n_{IR}(>M,r_{cl},z) = n_{IR}(>\lambda^{MF}_{IR}M,r_{ab},z),
\end{equation}
where the MF scaling factor $\lambda^{MF}_{IR}$ is given by
\begin{equation}
\lambda^{MF}_{IR}(r_{cl}/r_{ab})=
\lambda_{IR}(r_{cl}/r_{ab}) \frac{R(r_{cl})}{R(r_{ab})},
\end{equation}
or
\begin{equation}
\frac{M/L(r_{cl})}{M/L(r_{ab})}=
\frac{\lambda^{MF}_{IR}(r_{cl}/r_{ab})}{\lambda_{IR}(r_{cl}/r_{ab})}.
\end{equation}
Namely, the scale-dependence of the $M/L$ ratio is completely determined 
by the ratio between the MF and RF scaling factors of the IRAS 
$r_{cl}$-clusters. 

However, despite that we have the observed scaling factor 
$\lambda_{IR}(r_{cl}/r_{ab})$, we are in lack of the observed 
$\lambda^{MF}_{IR}(r_{cl}/r_{ab})$. This problem can be overcome by matching 
the IRAS galaxy traced mass field with popular models of CDM cosmogony.
With properly selected parameters, CDM cosmogony models
have been found to provide good approximation to many observational
properties of large scale structures, especially the mass function of
optical and X-ray clusters (e.g. Jing \& Fang 1994; Bahcall, Fan \& Cen 1997, 
and references therein). On the other hand, some individual clusters and 
superclusters from optical surveys have been identified from the IRAS data 
(Fisher et al. 1995; Webster et al. 1997). Therefore, it is reasonable
to assume that the mass field of CDM cosmogony model provides 
a description for the mass field traced by the IRAS galaxies. Thus,
the statistical properties of the simulated $r_{cl}$-clusters should be 
the same as that of the IRAS $r_{ab}$-clusters. We have then
\begin{equation}
\lambda(r_{cl}/r_{ab})=\lambda^{MF}_{IR}(r_{cl}/r_{ab}),
\end{equation}
i.e. the scaling factors of $n(>M,r_{cl})$ and $n_{IR}(>M,r_{cl})$ are the
same. Using eqs. (9) and (10) we have
\begin{equation}
\frac{M/L(r_{cl})}{M/L(r_{ab})}=
\frac{\lambda(r_{cl}/r_{ab})}{\lambda_{IR}(r_{cl}/r_{ab})}.
\end{equation}
Thus, the $M/L(r_{cl})$ ratio can be determined by the ratio between
the model-predicted $\lambda(r_{cl}/r_{ab})$ and the observed 
$\lambda_{IR}(r_{cl}/r_{ab})$.

The ratio $\lambda/\lambda_{IR}$ as a function of $r_{cl}/r_{ab}$ is 
plotted in Fig. 4. For LCDM and OCDM, the value of this ratio does not 
depart from 1 with a significance larger than 1-$\sigma$ in the entire 
range of scales from 3 to 24 $h^{-1}$ Mpc. In the case of SCDM, only at
$r_{cl}=$ 24 $h^{-1}$ Mpc, the ratio shows a difference from 1 with 2 
$\sigma$ level. Particularly, in the redshift shell [2500,5000] km s$^{-1}$,
the ratio is very flat for all the three models. Thus, for IRAS sample,
the $M/L$ ratio is basically flat on large scales from 3 to 24 $h^{-1}$ Mpc, 
i.e. 
\begin{equation}
\frac{M/L(r_{cl})}{M/L(r_{ab})} \simeq 1.
\end{equation}
This result is {\it weakly} dependent on specific parameters of models
of the CDM cosmogony. The $\chi^2$ goodness is larger than 98\% for
all the three CDM models.

Nevertheless, in Fig. 4, we can see a slight, but systematic increase 
of $M/L$ with scale. A linear fitting by 
$\log (\frac{M}{L}(r_{cl})/\frac{M}{L}(r_{ab}) )
\simeq a +b\log r_{cl}$ yields  $\chi^2$ goodness larger than 99.9\% for 
all the three models. The least-square fitted slope $b$ is found to be 
$0.19\pm0.17$ for OCDM, $0.21\pm0.17$ for LCDM, and $0.41\pm0.17$ for SCDM. 
That is, the $M/L$ ratio on scale $\sim 20 h^{-1}$ Mpc is probably larger 
than that of $r_{ab}$ by a factor about 1.7 for LCDM and OCDM. (or, a 
factor of 2.9 for SCDM.) These results for OCDM and LCDM are consistent with 
the $M/L$ ratio measurement of the Corona Borealis supercluster. 
Namely, $M/L$ scale-dependence given by IRAS galaxies is consistent with 
optical data.

\section{Conclusion}

The RF scaling factor of multi-scale identified $r_{cl}$-clusters is 
fundamentally important as a statistical measure for clustering  on
scales larger than clusters. Since these RFs depend only on the number of 
member galaxies of $r_{cl}$-clusters, they can be directly detected from 
a galaxy redshift survey regardless the mass determination of individual 
objects.  

The $M/L$ ratio of gravitationally clustering systems which consist of 
luminous objects and dark matter depends only on the scale of the system 
but not on total mass (Bahcall and Cen 1993). This is actually a scaling 
of cosmic gravitationally clustering. Therefore, the scale dependence of 
$M/L$ ratio is closely related to other scaling indices of these 
objects. We have shown that the scale behavior of $M/L$ is determined by the 
RF and MF scaling factors of $r_{cl}$-clusters.

By comparing the mass function scaling of simulated samples with the 
richness function scaling of IRAS 1.2 Jy galaxies, we conclude that the 
assumption of $M/L$-flatness is basically correct in the range from 
1.5 to 24 $h^{-1}$ Mpc. This conclusion is weakly dependent on the 
cosmological parameters we have considered.

\acknowledgements

We would like to thank an anonymous referee for a detailed report
which improves the presentation of the paper.

\newpage

\begin{figure}
\figcaption{Mass functions of clusters identified with radii
$r_{cl}=$ 1.5, 3.0, 6.0, 12 and 24 $h^{-1}$ Mpc for models of SCDM, LCDM
and OCDM at the epoch $z=0$. $n(>M,r_{cl})$ is the number density 
of $r_{cl}$-clusters with masses larger than $M$ in unit
of $h^3$Mpc$^{-3}$. $M$ is in unit of $h^{-1}$ M$_{\odot}$.}
\label{fig1}
\end{figure}

\begin{figure}
\figcaption{Richness functions of multi-scale identified 
$r_{cl}$-clusters from IRAS galaxies in the redshift shells 
of [2500,5000], [5000,7500] and  [7500,10000] km s$^{-1}$, with 
$r_{cl}= 1.5, 3.0, 6.0$ and $12 h^{-1}$ Mpc. $N_g$ is the number of 
member galaxies of the IRAS $r_{cl}$-clusters.
$n_{IR}(>N_g,r_{cl})$ is the number density of $r_{cl}$-clusters with 
number of member galaxies more than $N_g$.}
\label{fig2}
\end{figure}

\begin{figure}
\figcaption{Scaling factor $\lambda_{IR}(r_{cl}/r_{ab})$ as a function of
$r_{cl}$. The error bars come from the clusters identified from IRAS
galaxies. The triangle, circle and cross are, respectively for the redshift
shells [2500,5000], [5000,7500] and [7500,10000] km s$^{-1}$. The triangle
and cross data are horizontally shifted, respectively, by -0.01dex and
+ 0.01dex for clarity.}
\label{fig3}
\end{figure}

\begin{figure}
\figcaption{The ratio $\lambda(r_{cl}/r_{ab})/\lambda_{IR}(r_{cl}/r_{ab})$
or $(M/L(r_{cl}))/(M/L(r_{ab}))$ as a function of scale $r_{cl}$.
The symbols are the same as those in Fig.3.}
\label{fig4}
\end{figure}

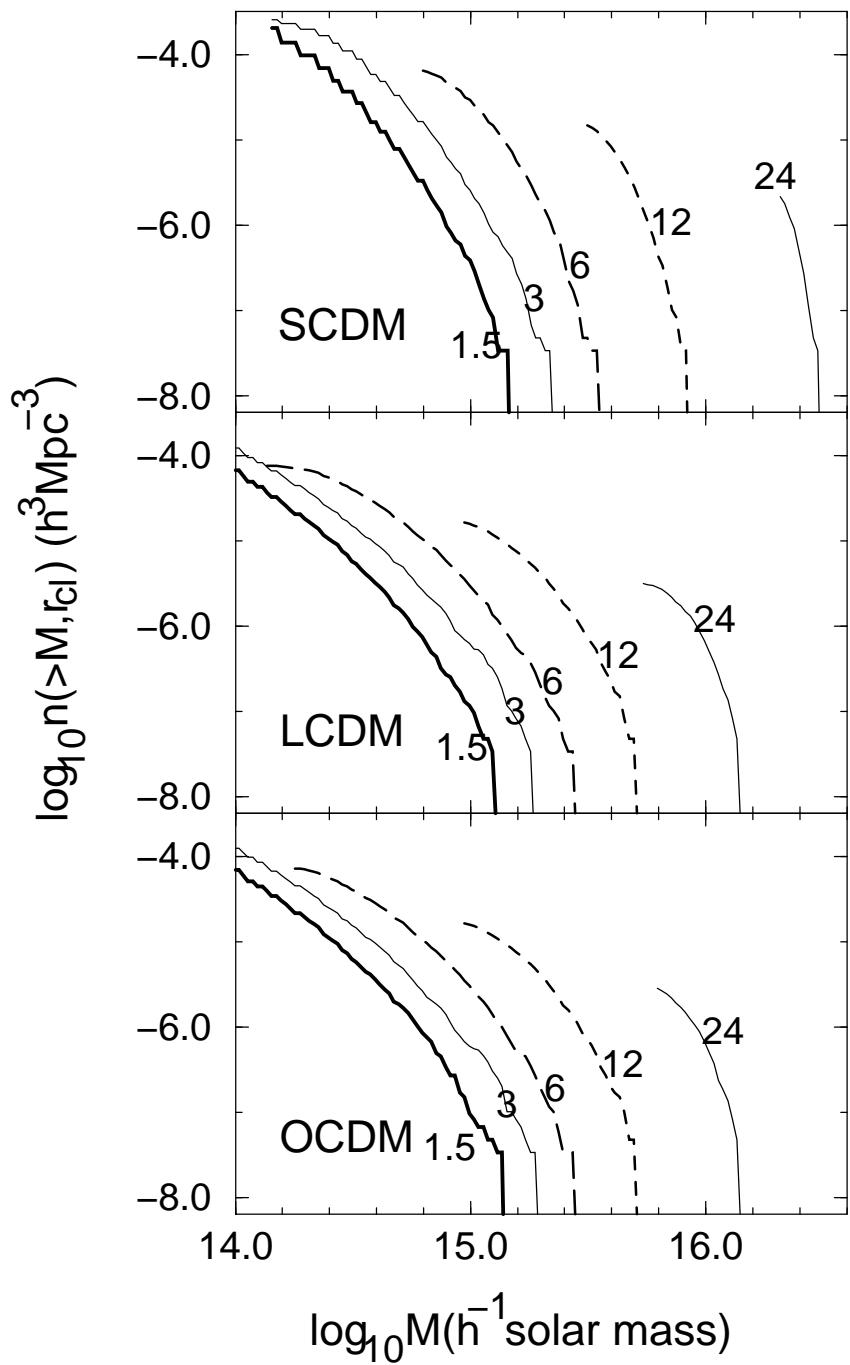
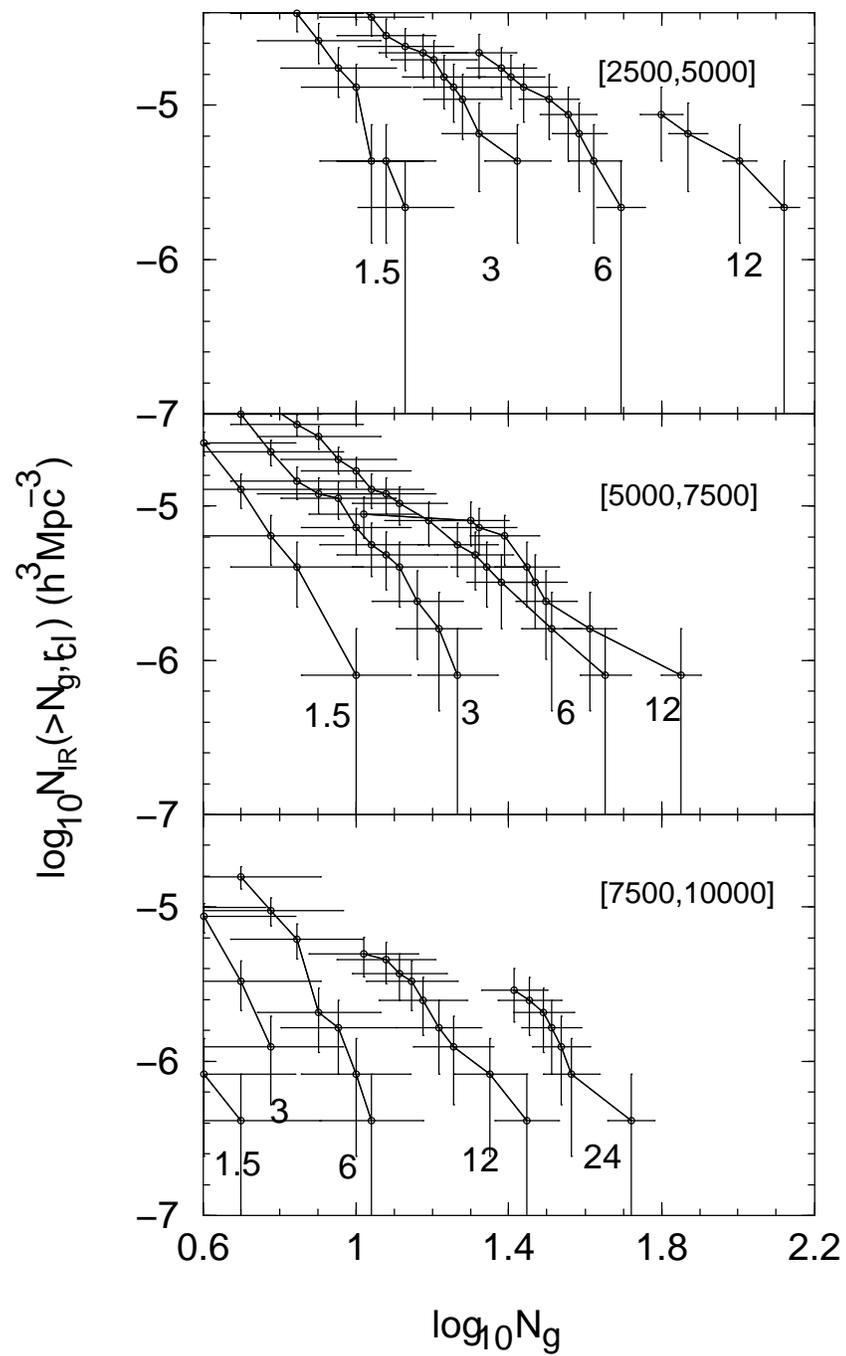
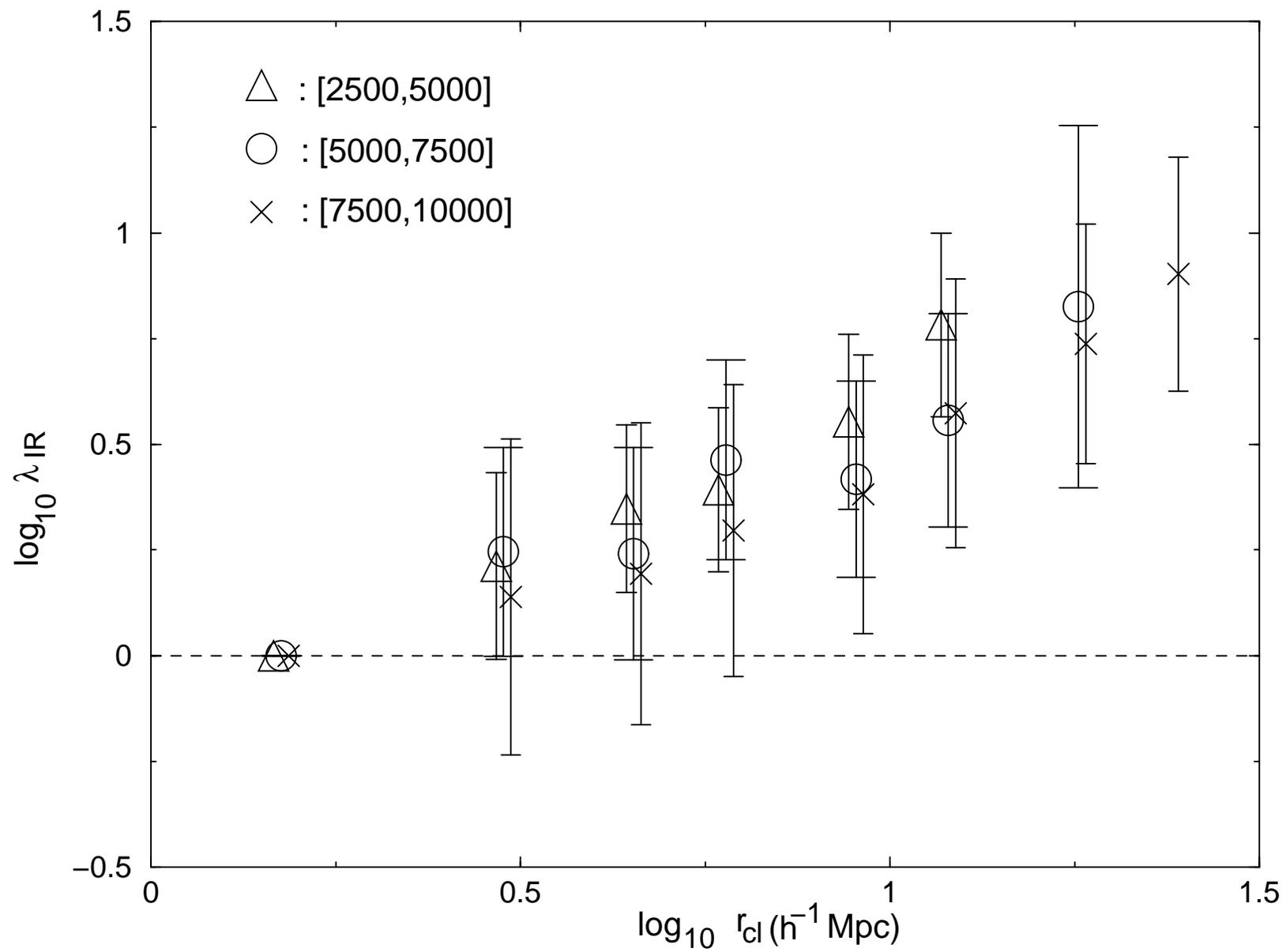
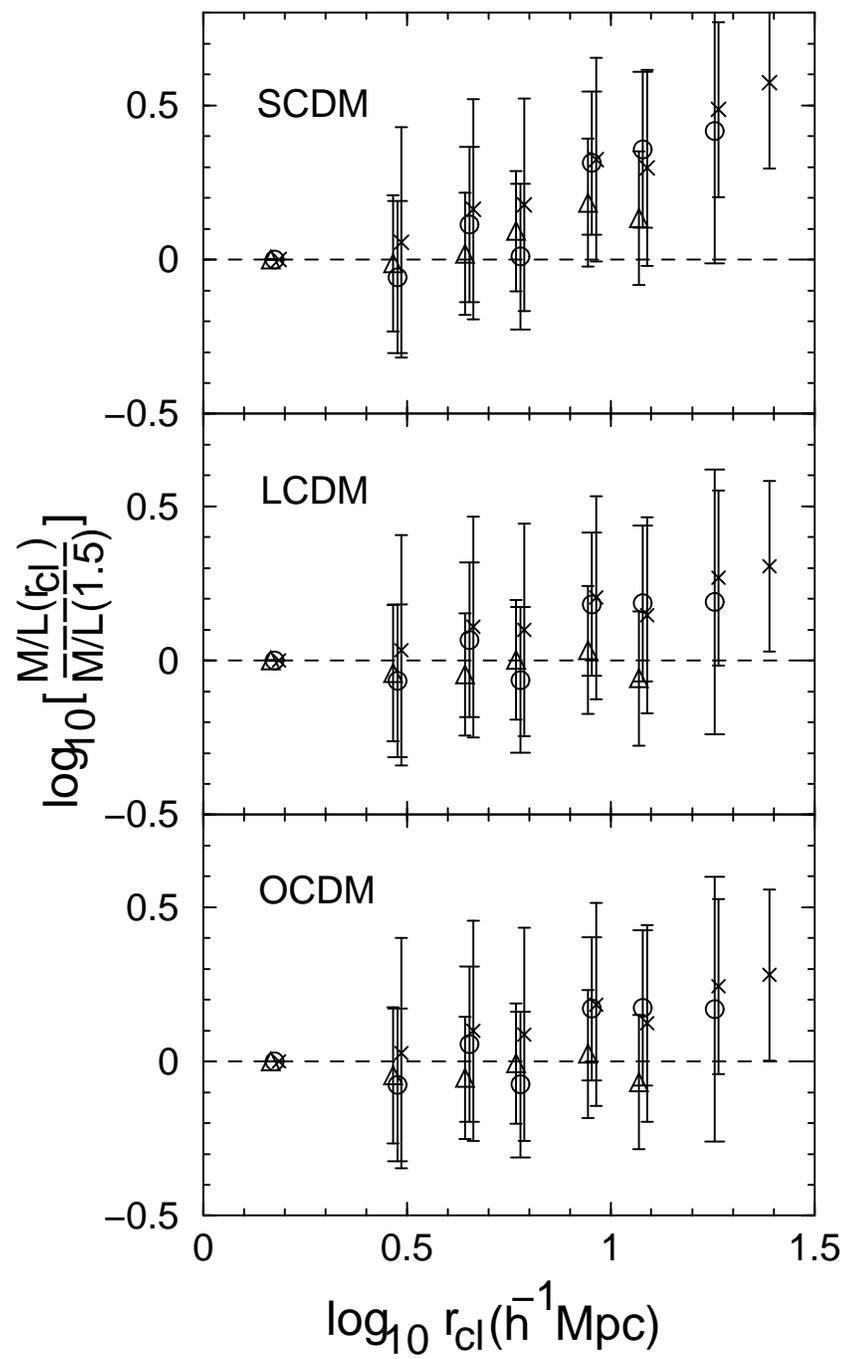
\end{document}